\documentclass[12pt]{iopart}
\usepackage{graphics,epsf,amsbsy,iopams} 
\usepackage[dvips]{graphicx}
\newcommand{\beq}{\vspace*{4mm} \begin{equation}}
\newcommand{\eeq}{\vspace*{4mm} \end{equation}}
\newcommand{\beqa}{\vspace*{4mm} \begin{eqnarray}}
\newcommand{\eeqa}{\vspace*{4mm} \end{eqnarray}}
\begin{document}

\title{Mean First Passage Time in Periodic Attractors}

\author{Avner Priel} 

\address{\small Department of Physics, University of Alberta, Edmonton, AB T6G2J1, Canada}

\ead{apriel@phys.ualberta.ca}

\begin{abstract}
The properties of the mean first passage time in a system characterized by multiple periodic attractors are studied. Using a transformation from a high dimensional space to 1D, the problem is reduced to a stochastic process along the path from the fixed point attractor to a saddle point located between two neighboring attractors.
It is found that the time to switch between attractors depends on the effective size of the attractors, $\tau$, the noise, $\epsilon$, and the potential difference between the attractor and an adjacent saddle point as:  $~T = {c \over \tau} \exp({\tau \over \epsilon} \Delta {\cal{U}})~$; the ratio between the sizes of the two attractors affects $\Delta {\cal{U}}$. 
The result is obtained analytically for small $\tau$ and confirmed by numerical simulations. 
Possible implications that may arise from the model and results are discussed.

\end{abstract}

\vspace{2pc}

\maketitle

\section{Introduction}

The problem of evaluating the mean first passage time (MFPT), has been investigated extensively in the context of chemical reactions, dynamical systems etc. \cite{moss,talkner_1,graham}. More recently this problem has been applied to the biological domain, e.g., kinesin walking on a microtubule \cite{kolomeisky}, DNA transport through membrane channels \cite{lakatos}, and the time to switch between protein folding states \cite{micheelsen}.

In this paper the focus is set on dynamical systems characterized by multiple periodic attractors. The system may be continuous, however the formal derivation is given for discrete time. The problem can be viewed as follows; consider the dynamics in a phase space governed by attractors, each represents a dynamic variable. The system evolves by changing the amplitudes of these variables. The system being in a particular attractor is reflected by the dominance of the associated variable in the current system's state. In the absence of noise, the system evolves into one of the available attractors, depending on the initial condition. However, when noise is added to the dynamics the system may escape from the basin of attraction. The MFPT is then defined as the mean time it takes to drive such a system from an attractor to an adjacent saddle point. A special case of this problem that arises in the context of a neural network model with a feedback loop has been treated previously. This family of model networks is in fact a system whose dynamics is influenced by the structure of the attractor space \cite{priel_98}. 
A geometrically-driven approach to the problem has been proposed in which the MFPT is calculated along a 1D path embedded in the higher dimensional attractor space. This path is a valley connecting the (periodic) fixed-point attractor and the metastable saddle point.
In periodic systems and the absence of noise, the coupled equations evolve into one of the mutually competing fixed points (f.p.'s) in which only the Fourier components representing this attractor has a non-vanishing coefficient, i.e.\ only one of the periodic orbits exists asymptotically. The addition of noise generates a perturbation in each of the coupled equations.
The perturbation can ``kick'' the system out of the vicinity of one stable f.p.\ so that it escapes to the other f.p. 
Our motivation is to quantify the mean time for such an event to occur. 

In this paper the derivation is extended to arbitrary asymmetric attractors that arises from a more general nonlinear dynamical system. An analytical result for the MFPT in the limit of weak noise and a weakly non-linear map is derived and confirmed numerically. The main reasons for taking these limits are as follows; a high level of noise may drive the system too fast from one basin of attraction to another, hence the actual dynamics of the system becomes less relevant. The reason for taking  weak non-linearity is two-fold; one is to facilitate the analytical derivation and allows for certain approximations; the other is that most of the characteristics are already revealed in this limit.

\section{The Model}

The family of systems analyzed here can be described by a set of coupled nonlinear recurrent equations,
generally given by the map:
\beq\label{simple_map}
x^i_{n+1} = f_i(X_n) + \xi_n 
\eeq 
\noindent
where $f(X)$ is a non-linear function, $x^i$ a component of $X$. The noise term, $\xi$, 
is a Gaussian additive noise distributed according to:

\beq\label{noise_dist}
\rho_{\epsilon} (\xi) = \frac{1}{(2 \pi \epsilon)^{1/2}} 
\exp - \frac{\xi^2}{2 \epsilon} \quad .
\eeq

Without loss of generality and to facilitate the derivations and the graphical presentation, we shall focus on the 2D case.  
In this case the map would be:
\beq\label{simple_2dmap}
\left\{ 
\begin{array}{ccc}
x_{n+1} & = & f_1(x_n,y_n) + \xi^1_n \\ \\
y_{n+1} & = & f_2(x_n,y_n) + \xi^2_n 

\end{array}
\right.
\eeq 
\noindent

Since the model deals with periodic systems, bounded nonlinear maps are of interest. In this case
each attractor may be characterized by a dominant Fourier component. Generally speaking, the 
periodic function may be expanded as an infinite series, however the demand for weak non-linearity 
allows for keeping only terms up to certain degree, e.g. 3'rd order. A typical such case might be:
\beq\label{typical_periodic_func}
A_i^{n+1}=g \left( \sum_m A_m^n \cos(\omega_m t) \right)
\eeq
\noindent
where $g()$ is the nonlinear function and $A$ is the amplitude of a Fourier component. The following 
map (\ref{asym_coupled_eq}) is obtained by assuming a bounded nonlinear function that has odd terms 
in the polynomial expansion. Taking only the 1'st and 3'rd order terms of $g$ leads to:
\beq\label{asym_coupled_eq}
\left\{ 
\begin{array}{ccc}
x_{n+1} = (1+\tau_1) x_n \left[ 1 - a x_n^2 - b y_n^2 \right] \\ \\
y_{n+1} = (1+\tau_2) y_n \left[ 1 - a y_n^2 - b x_n^2 \right] 
\end{array}
\right.
\eeq
where $0<a,b<1$ are constants and $|x_0|<1, |y_0|<1$.

The key point is our ability to identify a low dimensional discrete dynamics that describes the 
evolution of the system, i.e., the amplitude of the dominant Fourier components in the asymptotic
meta-stable state (w/o noise). 

Using the following variables rescaling:
\beq\label{var_rescale}
x=\sqrt{\tau_2} \tilde{x} \qquad y=\sqrt{\tau_1} \tilde{y} \quad ,
\eeq
\noindent
gives rise to the potential function
\beq\label{pot_func}
{\cal U}(\tilde{x},\tilde{y}) = - {\tilde{x}^2+\tilde{y}^2 \over 2} + 
{a \over 4} ({{\tau_2 \over \tau_1} \tilde{x}^4 + {\tau_1 \over \tau_2} \tilde{y}^4}) + b {\tilde{x}^2 \tilde{y}^2 \over 2} \quad ,
\eeq
\noindent
where the coupled equations are obtained via
\[
~ \tilde{x}_{n+1} = \tilde{x}_n - \tau_1 {\cal U}_x^{\prime} ~~~ , ~~~
~ \tilde{y}_{n+1} = \tilde{y}_n - \tau_2 {\cal U}_y^{\prime} ~.  
\]
\noindent
Keeping only terms with the small parameter $\tau$ to first order, the following 
transformed equations are obtained:

\beq\label{aymm_coupled_eq_trans}
\left\{ 
\begin{array}{ccc} \tilde{x}_{n+1} & = & \tilde{x}_n - \tau_1 \left(-\tilde{x}_n + a {\tau_2 \over \tau_1} \tilde{x}_n^3 + 
b \tilde{x}_n \tilde{y}_n^2 \right)  \\ \\
\tilde{y}_{n+1} & = & \tilde{y}_n - \tau_2 \left(-\tilde{y}_n + a {\tau_1 \over \tau_2} 
\tilde{y}_n^3 + b \tilde{y}_n \tilde{x}_n^2 \right) 
\end{array} 
\right. \quad ,
\eeq
\noindent

\begin{figure}[ht]
\includegraphics{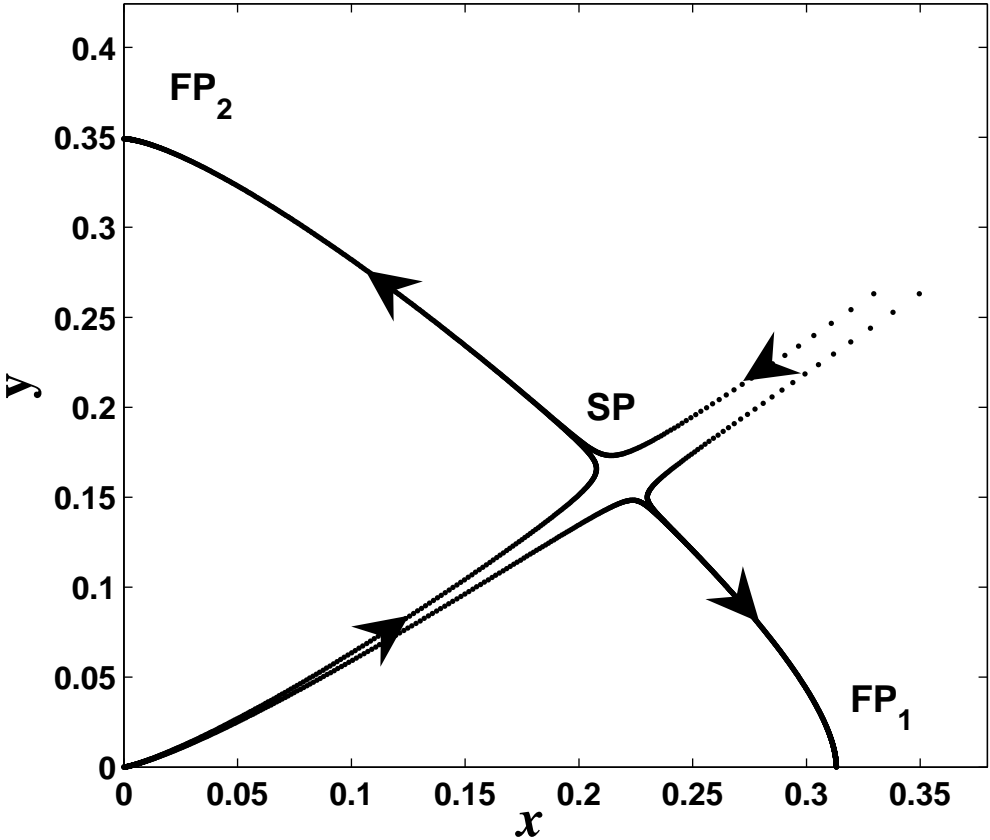}
\caption{Phase portrait of a 2D map with 
$\tau_1=0.02,~\tau_2=0.025,~a=0.2,~b=0.4$. 'SP' denotes a saddle point and 'FP' 
a fixed point. Arrows shows the direction of the flow.} 
\label{phase_portrait_asym}
\end{figure}

The fixed points of the dynamics (assuming $\tau_i \ll 1$) are given (for the rescaled variables) by 
\beq\label{asymm_fps}
\left[\pm \sqrt{\tau_1 \over a \tau_2}~,~0\right] ~,~ \left[0~,~\pm \sqrt{\tau_2 \over a \tau_1} \right]
 ~,~ \left[\pm \sqrt{{a \over a^2-b^2} ({\tau_1 \over \tau_2} -{ b \over a})} ~,~ 
\pm \sqrt{{a \over a^2-b^2} ({\tau_2 \over \tau_1} -{ b \over a})} \right] \quad ,
\eeq
\noindent
where the last (four) f.p.\ are in fact saddle points and the first (four)
are stable f.p. 
An example of the phase portrait in this case is given in figure \ref{phase_portrait_asym}. 

%
%
%  FIGURE 1 GOES HERE
%
%

\begin{figure}[ht]
\includegraphics[width=15cm]{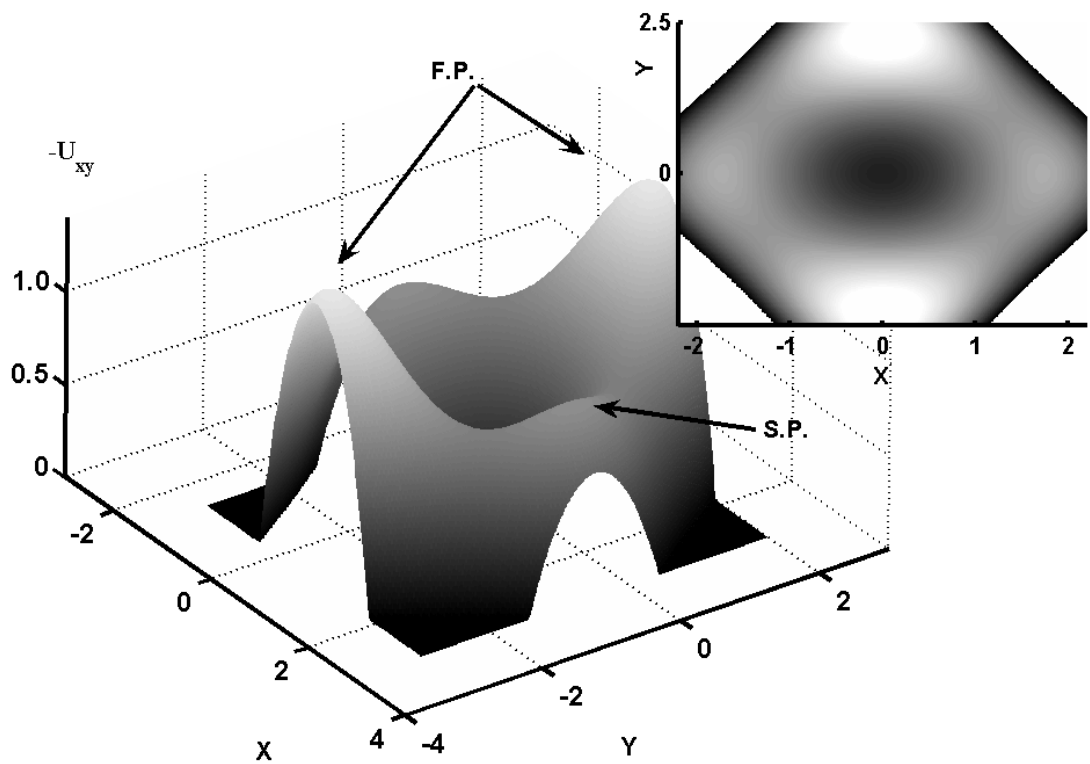}
\caption{Negative potential surface of the function given in equation \ref{pot_func}. 
The insert is the contour plot. The parameters used were 
${\tau_1 \over \tau_2} = 3/4, ~a=0.5,~b=0.25$. 'SP' denotes a saddle point and 'FP' 
a fixed point.} 
\label{fig_potential_surf}
\end{figure}

To visualize the potential surface an example is depicted in figure \ref{fig_potential_surf}; for clarity the
negative potential is shown - more 'red' means more negative potential (e.g., the zero F.P.\ is not stable).  
The variables shown are rescaled and the actual values of the potential are not important.  
It is apparent that there exists a path from a non-zero F.P.\  to adjacent S.P.\ via a valley 
(here a hill).

%
%
%  FIGURE 2 GOES HERE
%
%

As mentioned above, an important step in the derivation is the conjecture that the most probable
trajectory from a f.p.\ to one of its nearest s.p.\ defines the properties of the MFPT. 
In the next section the properties of the rescaled 1D noisy map with a potential given by equation 
\ref{pot_func} are derived. By fixing the initial and final points the path is unique, hence, 
the projection of near by trajectories on the path are treated and the noise term is rescaled appropriately.

\section{Escape from a meta-stable attractor}

To illustrate the idea behind the following derivation let us assume a scenario
where the initial condition is $y=0 ~,~ x=x^{\star}$, i.e.\ one of the
f.p.'s of the dynamics (refer to figure \ref{phase_portrait_asym}). Since the line connecting this f.p.\ 
and the saddle point is a valley (in the potential space), it may be assumed that the most probable 
escape route is along this line (or its mirror through the x-axis, i.e.\ the line
connecting the f.p.\ with the saddle point $( x_{sp}^{+}, y_{sp}^{-} )~$ ). 
This argument can be understood by rotating each noise term tangent and perpendicular to the 
path. The perpendicular term decays fast due to the restoring force; 
hence, the crucial step that the dynamics is mainly 1D is conjectured. 
It should be noted that this conjecture is not limited to our 2D example; rather, 
a set of attractors embedded in a higher dimensional space should exhibit the same 
property, namely, that the most probable escape route from the domain of attraction is 
through the valley connecting this attractor and a saddle-point residing between two adjacent 
f.p.'s. Therefore, with the assumption of weak noise and $\tau \ll 1$  the map can be reduced 
into one dimension, on that path; hence, a 1D noisy map is obtained:

\beq\label{1D_noisy_map}
s_{n+1} = s_n - \tau {\cal U'} (s_n) + \hat{\xi}_n \quad ,
\eeq
\noindent
where $s$ defines the path. The noise term now is the tangential
projection of the noise vector on the path. 
The explicit form of the path is not crucial as long as smoothness and monotonicity are guaranteed;
since it is assumed that there are no extremums between the f.p.\ and the s.p.\, this assumption holds.

This type of a 1D equation has been investigated  by several authors 
\cite{talkner_1,knessl,talkner_2} for the case of small non-linearity, namely, the class of 
map functions with the property that $f(s)$ deviates only weakly from the identity map:

\beq\label{map_class}
f(s) = s - \tau \frac{dU(s)}{ds} \quad , \qquad \tau \ll 1 \quad .
\eeq
\noindent
in analogy with our 1D map (equation \ref{1D_noisy_map}) . 

Assume that the process described in equation \ref{1D_noisy_map} is defined 
in $(- \infty , \infty )$ and let us define the random variable $\tilde{t}(s)$,
the first passage time from the interval 
$I= \left[ SP^{-}, SP^{+} \right] $ , by:
\beq\label{fpt}
\tilde{t} = \min \{ n : |s_n| \geq s_{sp}^{+} \} \quad ,
\eeq
\noindent
i.e.\ the first time the process hit one of the boundaries,
where $SP^{\pm}$ are the saddle points defined above, and 
$s_{sp}^{+}$ is the value of $s$ at the saddle point.
The MFPT, $t(s)$, starting from a point in $I$ is given by:
\beq\label{mfpt_def}
t(s) = < \tilde{t} (s) >  = E[~\tilde{t} ~|~ S_0 = s~] \quad .
\eeq
\noindent
It was shown (e.g., \cite{talkner_2}) that the MFPT can be written as
\beq\label{mfpt_int}
t(s) - 1 = \int_{I} P(z|s) t(z) dz \quad ,
\eeq
\noindent
where $~P(z|s)~$ denote the transition probability to go from $s_n=s$ to
$s_{n+1}=z$ in a single step.

In the next subsections the probability density function and the MFPT are derived.

\subsection{Probability density function}

Assume the noise terms, $\xi_n~$, are Gaussian distributed and mutual 
independent, hence, the stochastic process $s_n$ is a Markov process.  
The distribution is given by equation \ref{noise_dist}, with a rescaled amplitude 
$\epsilon$. The transition probability from $s_n=z$ to $s_{n+1}=s$ in one 
step is
\beq\label{trans_prob}
P(s|z)=\rho(s-f(z)) \quad ,
\eeq
\noindent
with $\rho(\xi)$ given by equation \ref{noise_dist}. 
The probability density, $Q_n(s)$, to find the system in $s$ after $n$ steps
evolves according to
\beq\label{evol_prob_density}
Q_{n+1}(s)=\int_{-\infty}^{\infty} P(s|z) Q_n(z) dz \quad . 
\eeq
\noindent
In a similar way, one can define the conditional probability $P_n(s|z)$ to get
from $z$ to $[s,s+ds]$ in $n$ steps.
The invariant probability density $Q(s)$ is the solution of equation 
\ref{evol_prob_density}. Plugging eq.\ \ref{trans_prob}, the invariant density
obey
\beq\label{inv_prob_density}
Q(s)=\int_{-\infty}^{\infty} P(s|z) Q(z) dz = \\
(2 \pi \epsilon)^{-1/2} \int \exp \left\{ -{1\over 2\epsilon}\left[s-z+\tau
{\cal U}^{\prime}(z)\right]^2 \right\} Q(z) dz
\quad . 
\eeq
\noindent
This integral can not be solved in general. In the case of weak non-linearity,
e.g.\ our map with $\tau \ll 1$, one can solve the integral approximately by 
setting $u=(2\epsilon)^{-1/2}\left(s-z+\tau{\cal U}^{\prime}(z)\right) $
and neglecting ${\cal O}(\tau^2)$ terms in the expansion of $z(u)$. 
Alternatively, one can assume a WKB-type solution (e.g., \cite{kubo73})
to solve equation \ref{inv_prob_density} for weak noise, 
$\epsilon \ll 1$, 
\beq\label{inv_prob_wkb}
Q(s)=N(s) \exp\left(-{\Phi(s) \over \epsilon}\right) \quad ,
\eeq
\noindent
where $\Phi(s)$ is a potential function and $N(s)$ is a normalization factor.
Inserting this type of solution back in equation \ref{inv_prob_density}, one obtains:
\beq\label{inv_prob_prefactor}
N(s)= (2 \pi \epsilon)^{-1/2} \int N(z) \exp \left[-{1\over 2\epsilon}
W^2(s,z) \right] dz \quad ,
\eeq
\noindent
where 
\beq\label{inv_prob_W}
W^2(s,z)=2\left[\Phi(z)-\Phi(s)\right]+\left[s-f(z)\right]^2 \quad ,
\eeq
\noindent
whereas in our model $~f(z)=z-\tau{\cal U}^{\prime}(z)$. The demand for (semi) 
positivity of $W^2$ implies that the potential $\Phi(s)$ is a Lyapunov 
function of the noise-free dynamics. This is evident by setting $s=f(z)$ 
in equation \ref{inv_prob_W}. To obtain an equation for the potential $\Phi(s)$
one can derive a transformation from the old variable, $z$, to the new one,
$W$. It is sufficient and necessary that 
$~\partial W(s,z) / \partial z \neq 0~$ to write
\beq\label{inv_prob_var_trans}
N(s) = (2 \pi \epsilon)^{-1/2} \int N(z(W,s)) 
\left[ \partial W(s,z) / \partial z |_{z=z(W,s)} \right]^{-1}
\exp \left[-{1\over 2\epsilon} W^2 \right] dW \quad ,
\eeq 
\noindent
where $z(W,s)$ is a solution of equation \ref{inv_prob_W}. Assuming 
$~N(z(y,s)) \left[ \partial W / \partial z |_{z=z(W,s)} \right]^{-1}~$ 
is smooth in a neighborhood of ${\cal O}(\epsilon^{1/2})$ around $W=0$,
the integral can be approximated for weak noise by
\beq
N(s)=N(z(y=0,s)) \left[ \partial W / \partial z |_{z=z(W=0,s)} \right]^{-1}
\quad .
\eeq

Assuming the variable transformation remains valid 
($~\partial W / \partial z \neq 0~$) for the case 
$~\partial W^2(s,z) / \partial z = 0~$, $W^2~$ must vanish as well. Therefore,
differentiating \ref{inv_prob_W} leads to an equation for $s$. Plugging
$W^2=0$ and the equation for $s$ back to equation \ref{inv_prob_W}, 
the desired relation for $\Phi(s)$ is obtained:
\beq\label{inv_prob_potential}
\Phi(z)-\Phi\left({\Phi^{\prime}(z) \over 1-\tau {\cal U}^{\prime\prime}(z)} +
z-\tau {\cal U}^{\prime}(z) \right) + 
{1 \over 2} \left( {\Phi^{\prime}(z) \over 1-\tau {\cal U}^{\prime\prime}(z)}\right)^2=0 
\quad ,
\eeq
\noindent
where the map $f$ has been written explicitly and ${\cal U}^{\prime\prime}(z)$
is a second order derivative w.r.t.\ $z$.
The solution of this equation for small non-linearity, $\tau \ll 1$, gives 
finally
\beq\label{inv_prob_final_potential}
Q(s)=N \exp {-2 \tau {\cal U}(s) \over \epsilon} \quad ,
\eeq
\noindent
where the prefactor $N$ is constant up to corrections of ${\cal O}(\tau^2)$.\\

\subsection{The MFPT}

Turning now to the analysis of the MFPT, consider the random variable 
$\tilde t(s)$ - the random time to leave the domain of attraction, $I$, 
for the first time, starting from a point $s \in I$. 
The probability that after $n$ steps the process does not leave the domain $I$
is given by
\beq\label{w(n|s)}
Q(n|s) = \int_I P_n(z|s) dz \quad ,
\eeq
\noindent
where $P_n(z|s)$ is the probability to get from $s$ to $z$ in exactly $n$ 
steps, obeying a similar equation as \ref{evol_prob_density}
\beq\label{prob_backward_eq}
P_{n+1}(z|s)=\int_I P_n(z|y) P_1(y|s) dy \quad ,
\eeq
\noindent
where this recursive relation uses the one-step probability $P_1$.
Since $Q(n|s)$ is a decreasing function of the discrete time $n$,
the probability that an exit occurs exactly after $n$ steps is simply
$~Q(n|s)-Q(n+1|s)~$, hence, the first moment of our random variable,
$\tilde t(s)$, is 
\beq\label{mfpt_eq1}
t(s)=\langle \tilde t(s) \rangle = \sum_{n=0}^{\infty} 
(n+1) (Q(n|s)-Q(n+1|s)) = \sum_{n=0}^{\infty} Q(n|s) \quad .
\eeq
\noindent
Summing equation \ref{evol_prob_density} over $n$ and taking the integral inside 
the domain $I$ only, one obtains:
\beq\label{mfpt_eq2}
\sum_{n=0}^{\infty} Q(n+1|s)= \int_I \sum_{n=0}^{\infty} Q(n|s) P_1(z|s) dz
\quad .
\eeq
\noindent
On noting that the l.h.s.\ equals $~t(s)-1~$ ($~Q(0|s)=1 ~\forall ~s \in I~$)
this equation becomes equation \ref{mfpt_int}. 
Multiplying equation \ref{mfpt_int} by the invariant probability density $Q(s)$ 
and integrating over the domain as described by equation \ref{fpt}, i.e.,
$~I= \left[ SP^{-}, SP^{+} \right]~$, it is found that:
\beq\label{mfpt_eq3} 
\int_{SP^{-}}^{SP^{+}} Q(s) ds = 
\left( \int_{-\infty}^{SP^{-}} ds + \int_{SP^{+}}^{\infty} ds \right) Q(s)
\int_{SP^{-}}^{SP^{+}} P_1(z|s) t(z) dz \quad . 
\eeq
\noindent

Under the assumption of weak noise $\epsilon \ll 1 $, the function
$t(s)$ is nearly constant inside the domain of attraction. Fluctuations 
occur mainly near the boundary. The reason is that only close to the 
boundary may one have a finite probability to jump over the 
boundary in a small number of steps. Therefore, it was suggested
(see \cite{knessl,talkner_2}) that this function can be written as a product of 
a constant value, and a boundary layer function:
\beq\label{boundary_function}
t(s) = T \tilde{h}(s) ~~~~,~~ \tilde{h}(s^{\star}) = 1 \quad ,
\eeq
\noindent
where $s^{\star}$ is the f.p.
The boundary layer extends a distance of order ${\epsilon}^{1/2}$ around 
$s=s_{sp}^{+}$, and the scaled boundary layer function $h(s)$ may be written as:
$h(s)=\tilde{h}((2\epsilon)^{1/2}s)$. 

Using the boundary layer function $\tilde{h}$ (equation \ref{boundary_function})  
to account for the fact that for weak noise t(s) is almost constant inside the
domain, except near the boundaries, MFPT (starting from the f.p.) is given by:
\beq\label{mfpt_eq4}
T^{-1} ={\left(\int_{-\infty}^{SP^{-}}ds+\int_{SP^{+}}^{\infty}ds \right) Q(s)
\int_{SP^{-}}^{SP^{+}} P_1(z|s) \tilde{h}(z) dz \over 
\int_{SP^{-}}^{SP^{+}} Q(s) ds } \quad ,
\eeq
\noindent
where the first integral in the numerator is the inverse MFPT from the negative
exit and the second from the positive. Since the saddle points are symmetric
w.r.t.\ the f.p.\  the integrals equal. The invariant density is peaked
around the stable f.p., hence, the steepest descent approximation may be used
for both the numerator and the denominator of equation \ref{mfpt_eq4}. 
Expanding the map near the f.p., 
\beq\label{gauss_expand}
f(z^{\star}+\Delta z) \approx z^{\star}+\Delta z f^{\prime}(z^{\star})=
z^{\star}+\Delta z (1-\tau {\cal U}^{\prime\prime}(z^{\star})) \quad ,
\eeq
\noindent
where $z^{\star}$ denotes the f.p., and using the Gaussian approximation of 
equation \ref{inv_prob_density}, the denominator of equation \ref{mfpt_eq4} becomes
\beq\label{mfpt_eq4_denom}
\int_{SP^{-}}^{SP^{+}} Q(s) ds \approx Q(FP) 
\left[{ \pi \epsilon \over \tau {\cal U}^{\prime\prime} (FP)} \right]^{1/2}
\quad ,
\eeq
\noindent
where the weak non-linearity assumption, $\tau \ll 1$, is used. The numerator
is evaluated in a similar manner yielding 
\beq\label{mfpt_eq4_numer}
\left(\int_{-\infty}^{SP^{-}}ds+\int_{SP^{+}}^{\infty}ds \right) Q(s)
\int_{SP^{-}}^{SP^{+}} P_1(z|s) \tilde{h}(z) dz \approx 
\epsilon^{1/2} G(\tau) Q(SP) \quad ,
\eeq
\noindent
with $~G(\tau)=({\tau \over 2 \pi})^{1/2} + O(\tau)~$. 

Combining the terms, the main result is obtained:
\beq\label{mfpt_main_result}
T = {{\cal C} \over \tau} {Q(SP) \over Q(FP)} = {{\cal C} \over \tau}
\exp {2 \tau \over \epsilon} \left( {\cal U}(SP) - {\cal U}(FP) \right) \quad ,
\eeq
\noindent
where ${\cal C}$ is a constant. 
Note that the time $T$ is rescaled w.r.t.\ a cycle of the attractor. 

Plugging the general form of the f.p.'s of the dynamics (equation \ref{asymm_fps}) in equation \ref{pot_func}, 
an explicit expression for the potential terms in equation \ref{mfpt_main_result} 
as a function of the variables of the model, $\tau_1, \tau_2, a, b$, is obtained as follows:
\beq\label{potential_explicit_result}
 {\cal U}(SP) = - { a \tau_{12}^2 + a - 2b \tau_{12} \over 4 \tau_{12} (a^2-b^2)} \quad
{\cal U}(FP) = - {\tau_{12} \over 4 a}  ~~ ( - { 1 \over 4 a \tau_{12}} )
\eeq
\noindent
where $\tau_{12}=\tau_1 / \tau_2$; the parentheses on the r.h.s.\ are given for the 
second f.p.\ Note that $\tau_1, \tau_2$ are given now w.r.t.\ a common variable $\tau$, e.g., $\tau_2=\tau$
and $\tau_1=\tau_{12} \tau$. The special case $\tau_1=\tau_2=\tau$ reduces to
$ \left( {\cal U}(SP) - {\cal U}(FP) \right)= (b-a)/(4a(a+b)) $, and for the choice $a=1/4, b=1/2$ used below for simulations,
$ \Delta {\cal U} = 1/3 $.

\subsection{Numerical simulations}

To confirm the theoretical results extensive numerical simulations of the model were performed. 
Equations \ref{aymm_coupled_eq_trans} were used to evaluate the statistical properties of the MFPT.
For each trial, a line that passes through the s.p.\ and perpendicular to the 
line connecting the f.p.\ and the s.p.\ was constructed. The time to hit this line starting from the f.p.\ was
collected in $500-1000$ trials for each choice of the parameters. 
The model parameters used in the figure were $a=1/4, b=1/2$ (although other values 
were tested as well). Typically the attractor and noise parameters were chosen such that 
$\tau / \epsilon \in (2..7)$, and $0.8 \leq \tau_{12} \leq 1.2$.

\begin{figure}[ht]

\includegraphics{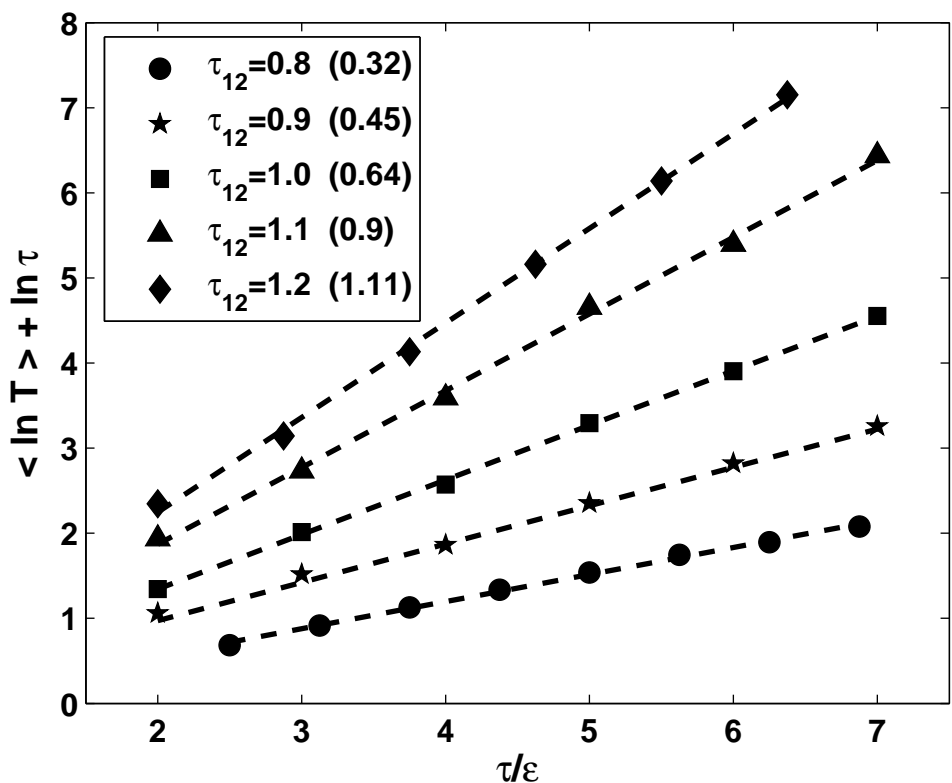}
\caption{Scaling of the average logarithm of the escape time in the
2D model. The dashed lines are the linear regression of the corresponding data points.
The standard deviation of each point is roughly the same as the symbol, hence omitted for 
clarity. The legend shows for each $ \tau_{12}=\tau_1 / \tau_2$ the estimated slope in parentheses,
see also table \ref{tab_predictedSlopes}.}
\label{sim_2d_fig}
\end{figure}

To demonstrate the scaling properties of the results, equations \ref{mfpt_main_result} - \ref{potential_explicit_result},
the logarithm of the MFPT is evaluated and the following quantity is depicted in the next figure:
\[
< \ln T > + \ln \tau ~ \propto ~ \tau / \epsilon
\]
\noindent
The slope of each series of simulations corresponds to a specific choice of $ \tau_{12}, \tau, \epsilon$.
The simulation results shown in figure \ref{sim_2d_fig} are in a good agreement with the 
predicted values calculated using equation \ref{potential_explicit_result}, see table \ref{tab_predictedSlopes}.

\begin{table*}[ht]
	\centering
		\begin{tabular}{|c|c|c|} \hline
		   $ \boldsymbol{\huge \tau}_{\bf 12} $ & $ \mathbf{ 2 \left( \boldsymbol{\cal U}(SP) - \boldsymbol{\cal U}(FP) \right) }$ &
		   \textbf{Simulation} \\ \hline \hline
			0.8 & 0.3 & 0.32 $\pm$ 0.02 \\ \hline 
			0.9 & 0.47 & 0.45 $\pm$ 0.03  \\ \hline
			1.0 & 0.66 & 0.64 $\pm$ 0.03 \\ \hline
			1.1 & 0.88 & 0.90 $\pm$ 0.03 \\ \hline
			1.2 & 1.09 & 1.11 $\pm$ 0.03 \\ \hline 
		\end{tabular}
	\caption{Predicted vs.\ simulated potential difference values}
	\label{tab_predictedSlopes}
\end{table*}

%
%
%  FIGURE 3 GOES HERE
%
%

\section{Discussion}

The theory for the mean first passage time to escape
a periodic attractor defined by the particular fixed point
of the dynamic equations was developed. 
To facilitate the analytical investigation, proper variables transformation were 
identified that decompose the potential function of the system.
The reduced variables are closely related to the amplitude of the solution. 
The noiseless system relaxes to one of the stable non-zero solution (above bifurcation). 
Adding noise to the dynamics perturbs this solution and enables possible escape.

One of the key points in the solution is the conjecture that the most probable escape route
is essentially one dimensional along the path from the attractor to one of the adjacent 
saddle points. This enabled us to reduce the dimensionality of the dynamics into 
a 1D process. Taking the limit of small noise and not too far from the bifurcation 
the theory developed for noise driven discrete dynamical systems has been applied. 
The results resemble those obtained in systems with potential barrier 
undergoing a tunneling in the sense that the escape time has a polynomial 
prefactor and a leading exponential term. 

Simulations of the system with two variables have
shown that the theory developed, and especially the reduction to a 1D flow, are
applicable and provide a good prediction of the exponential term, as well as the 
polynomial prefactor.

It should be noted that the results are applicable to non periodic systems as well, as long as
one can transform the dynamic equations to a form similar to equations \ref{asym_coupled_eq}, i.e., 
up to third order. Also, systems with higher order terms may exhibit a very similar behavior assuming
these terms do not dominate the solutions and the discussion is restricted to solutions not too far from bifurcation, i.e.\, $\tau \ll 1$.

The implications that arise from this results may have practical application in fields such as
optimization, minimization problems, numerical analysis, parameter estimation in 
high dimensional complex systems (e.g., neural networks) and more. 
The common ground to these problems is the existence of (usually) high dimensional attractor space 
in which some dynamics is imposed, e.g., obtaining a minimal energy solution to optimization problem via some iterative algorithm. This algorithm evolves the solution preferably to a global minimum, however it may get trapped in local minima. To escape these attractors, one occasionally perturbs the solution to escape from the solution etc. It is of practical importance to assess the time to escape (affect simulation time) and possibly the required amplitude and duration of perturbation.
All these issues are left for future research. In this context it is interesting to mention a work by Baronchelli and Loreto 
\cite{baronchelli} on mean first passage time in graphs, where the focus is on the complexity of evaluating the MFPT 
on a particular node of a graph.

\ack
The author would like to thank Jack Tuszynski and Daniel ben-Avraham for 
reviewing the final version of this manuscript.

\newpage

\section*{Figures}

\end{document}